\documentclass[useAMS,usenatbib,usegraphicx,]{mn2e}
\usepackage{epstopdf}

\title[The South-West Cloud]{Major Substructure in the M31 Outer Halo: the South-West Cloud\thanks{This work was based on observations obtained with the MegaPrime/MegaCam, a joint project of CFHT and CEA/DAPNIA, at the Canada-France-Hawaii Telescope (CFHT) which is operated by the National Research Council (NRC) of Canada, the Institute National des Sciences de l'Univers of the Centre National de la Recherche Scientifique of France and the University of Hawaii.}}
\author[N. F. Bate et al.]{N. F. Bate$^{1}$\thanks{E-mail:
nbate@sydney.edu.au (NFB); geraint.lewis@sydney.edu.au (GFL)}, A. R. Conn$^{1}$, B. McMonigal$^{1}$, G. F. Lewis$^{1}$, N. F. Martin$^{2, 3}$, \newauthor A. W. McConnachie$^{4}$, J. Veljanoski$^{5}$, A. D. Mackey$^{6}$, A. M. N. Ferguson$^{5}$, \newauthor R. A. Ibata$^{2}$, M. J. Irwin$^{7}$, M. Fardal$^{8}$, A. P. Huxor$^{9}$, and A. Babul$^{10}$ \\
$^{1}$Sydney Institute for Astronomy, School of Physics, A28, University of Sydney, Sydney, NSW 2006, Australia\\
$^{2}$Observatoire Astronomique, Universite de Strasbourg, CNRS, F-67000 Strasbourg, France\\
$^{3}$Max-Planck-Institut f\"{u}r Astronomie, K\"{o}nigstuhl 17, D-69117 Heidelberg, Germany\\
$^{4}$NRC Herzberg Institute of Astrophysics, 5071 West Saanich Road, Victoria, British Columbia V9E 2E7, Canada\\
$^{5}$Institute for Astronomy, University of Edinburgh, Royal Observatory, Blackford Hill, Edinburgh, EH9 3HJ, UK\\
$^{6}$RSAA, The Australian National University, Mount Stromlo Observatory, Cotter Road, Weston Creek, ACT 2611, Australia\\
$^{7}$Institute of Astronomy, Madingley Road, University of Cambridge, CB3 0HA, UK\\
$^{8}$University of Massachusetts, Department of Astronomy, LGRT 619-E, 710 N. Pleasant Street, Amherst, Massachusetts, 01003-9305, USA\\
$^{9}$Astronomisches Rechen-Institut, Universit\"{a}t Heidelberg, M\"{o}nchhofstra\ss e 12-14, 69120 Heidelberg, Germany\\
$^{10}$Department of Physics \& Astronomy, University of Victoria, Elliott Building, 3800 Finnerty Rd, Victoria, BC, V8P 1A1, Canada\\}
\begin{document}

\date{Accepted 2013 November 3; Received 2013 November 3; in original form 2013 September 16}

\pagerange{\pageref{firstpage}--\pageref{lastpage}} \pubyear{2013}

\maketitle

\label{firstpage}

\begin{abstract}
We undertake the first detailed analysis of the stellar population and spatial properties of a diffuse substructure in the outer halo of M31. The South-West Cloud lies at a projected distance of $\sim100$~kpc from the centre of M31, and extends for at least $\sim50$~kpc in projection. We use Pan-Andromeda Archaeological Survey photometry of red giant branch stars to determine a distance to the South-West Cloud of $793^{+45}_{-45}$~kpc. The metallicity of the cloud is found to be $[Fe/H] = -1.3\pm0.1$. This is consistent with the coincident globular clusters PAndAS-7 and PAndAS-8, which have metallicities determined using an independent technique of $[Fe/H] = -1.35\pm0.15$. We measure a brightness for the Cloud of $M_V= -12.1$ mag; this is $\sim75$ per cent of the luminosity implied by the luminosity-metallicity relation. Under the assumption that the South-West Cloud is the visible remnant of an accreted dwarf satellite, this suggests that the progenitor object was amongst M31's brightest dwarf galaxies prior to disruption. 
\end{abstract}


\section{Introduction}
According to standard $\Lambda$CDM theory, accretion history is imprinted on a galaxy's stellar halo. Long dynamical timescales in the outer halo ensure that accreted systems dissolve slowly, offering a window into galaxy assembly over an appreciable period of cosmic history. Numerous examples of dwarf galaxies and globular clusters undergoing tidal interactions with the Milky Way (MW) now exist. The prototypical example is the Sagittarius dwarf galaxy. It was serendipitously discovered by \citet*{ibata+94} in a study of the Galactic bulge \citep{ibata+95}, and subsequently shown to possess a large stellar stream \citep{mateo+96} that extends across nearly the entire sky (e.g. \citealt{majewski+03}). More recently, the globular cluster Palomar 5 has been shown to possess a well-defined stream (e.g. \citealt{odenkirchen+01}; \citealt{odenkirchen+03}; \citealt{grillmair+06}). Several further features have been found in the SDSS which do not have any clear progenitors: for example, the Orphan Stream \citep{belokurov+07} and the overdensity associated with Bo\"{o}tes III \citep{carlin+09}.

Many of these substructures have low surface brightness ($V > 30$~mags/sq. arcsec; \citealt{bullock+05}). We can therefore only obtain information on resolved stellar content in extremely nearby galaxies. Andromeda (M31) is an ideal target for such studies, because it offers a global panorama unavailable in the Milky Way. This view of M31 has already revealed some surprising patterns, such as globular clusters coincident with stellar substructures \citep{mackey+10}, and coherence in satellite motions (\citealt{ibata+13}; \citealt{conn+13}) that recall the Vast Polar Structure in the Milky Way \citep*{pawlowski+12}. Large coherent structures such as these are difficult to reproduce in standard $\Lambda$CDM, and their existence has been used to argue for alternate theories of gravity (e.g. \citealt{kroupa12}; \citealt*{pawlowski+13}).

Studies of substructure in the stellar halo of M31 have largely focussed on dwarf galaxies (e.g. \citealt{martin+06}; \citealt{martin+09}; \citealt{richardson+11}; \citealt{collins+13}; \citealt{martin+13}) and globular clusters (e.g. \citealt{mackey+10}; \citealt{huxor+11}; \citealt{mackey+13}; \citealt{veljanoski+13}), in part because they are comparatively easy to isolate from the stellar background. In addition to the discovery of the vast co-rotating plane of dwarfs, a recent  kinematic study of 18 dwarf spheroidals in the M31 system has show that the majority are dark matter-dominated objects, obeying similar mass-size and metallicity-luminosity scalings to MW dwarfs \citep{collins+13}.

The globular cluster (GC) population of M31 is interesting for a variety of reasons. It appears to be considerably richer than the MW, consisting of well over 500 confirmed globular clusters (see the Revised Bologna Catalogue, \citealt{galleti+04}), including almost a hundred outer halo objects (\citealt{veljanoski+13} and references therein; Huxor et al. 2013 in prep). It is particularly interesting that the M31 outer halo GCs preferentially lie on stellar streams and other substructures \citep{mackey+10}. The statistical probability for such chance alignment is less than 1 per cent \citep{mackey+10}. Kinematic coherence amongst GCs projected on such substructures further implies that they are associated \citep{veljanoski+13}, which points to an accretion origin for these objects \citep{searle+zinn78}.

Studies of diffuse stellar substructures in M31 have tended to focus on inner halo features, such as the north eastern shelf, the northern spur, the G1 clump, and the giant stellar stream (\citealt{ferguson+05}; \citealt{richardson+08}; \citealt{faria+07}). The giant stellar stream, most prominent of numerous streams in the stellar halo of M31, has received the most detailed analysis. It was first detected by \citet{ibata+01}, and extends approximately 60 kpc in projection south-east of M31 (\citealt{ferguson+02}; \citealt{mcconnachie+03}; \citealt{ferguson+05}). \citet{fardal+06} used an analytic bulge-disc-halo model for M31 \citep{geehan+06} to construct test-particle orbits that matched the stream. They used these results to estimate the mass of the progenitor to the stream, as well as associate a number of other streams and shelves in the M31 stellar halo with the progenitor of the stream (\citealt{fardal+07}; \citealt{fardal+13}). 

The Pan-Andromeda Archaeological Survey (PAndAS; \citealt{mcconnachie+09}) revealed a wealth of substructure throughout the stellar halo of M31, extending out to very large radii. The giant stellar stream is the most luminous of these substructures, but several other streams and features are observed, pointing towards an active accretion history for M31. Substructures in the outer halo, where dynamical timescales are long, offer a unique opportunity to study an older phase of accretion. So far, this region has remained unexplored.

\begin{figure*}	
	\includegraphics[width=110mm, angle=270]{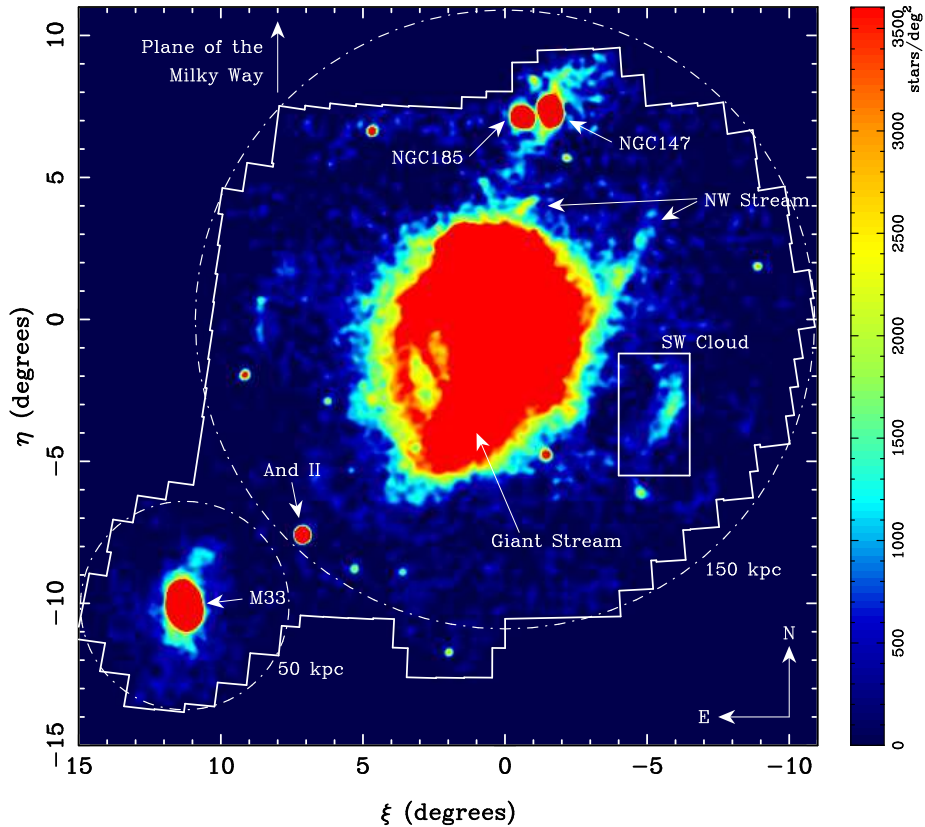}
  \caption{Stellar density map of PAndAS stars with de-reddened colours and magnitudes consistent with metal-poor red giant branch populations at the distance of M31. The map consists of $0.05\degr \times 0.05\degr$ pixels, and is smoothed using a Gaussian with $0.1\degr$ dispersion. It is displayed in tangent-plane projection centred on M31, with linear scaling chosen to highlight the location of the South-West Cloud. This map has been foreground contamination-subtracted, as described in Section \ref{sub:contamination}. Some features in the environment around M31 are marked; for a detailed schematic of prominent substructures see Figure 1 in \citet{lewis+13}.}
  \label{fig:fullmap}
\end{figure*}

In this paper we focus on one such outer-halo substructure: the South-West Cloud. It is a diffuse stellar overdensity with an estimated mass of $~10^7M_\odot$ \citep{lewis+13} that lies approximately 100~kpc south-west of M31 (see Figure \ref{fig:fullmap}). First observed in PAndAS \citep{mcconnachie+09}, this feature is coherent over approximately 50~kpc. We focus on its properties -- distance, metallicity, brightness -- and compare them with the Andromeda dwarf spheroidal population.

It is unclear whether the South-West Cloud is part of a stream in its own right, or a shell thrown off in some other interaction (\citealt{fardal+07}; \citealt{lewis+13}). It overlaps a large spur of HI gas, extending $\sim65$~kpc from the centre of M31. It is also spatially coincident with three globular clusters discovered in the PAndAS dataset, PAndAS-7 and PAndAS-8 (PA-7 and PA-8, \citealt{mackey+13}) and PAndAS-14 (PA-14, \citealt{veljanoski+13}). PA-7 and PA-8 are $[Fe/H]=-1.35\pm0.15$ globular clusters with unusually red horizontal branch morphologies, suggesting that they are at least $\sim2$ Gyr younger than the oldest MW globular clusters. 

The three globular clusters have similar radial velocities (PA-7: $-433\pm8$~km/s; PA-8: $-411\pm4$~km/s; PA-14: $-363\pm9$~km/s), well separated from the M31 systemic velocity (\citealt{mackey+13}; J. Veljanoski priv. comm.). These velocities are consistent with that of the \mbox{H\,{\sc i}} gas spur, although it should be noted that its errors are large ($-470\pm200$~km/s, \citealt{lewis+13}). The systemic radial velocity of the South-West Cloud is currently unknown; should the association of PA-7, PA-8 and PA-14 with the South-West Cloud be confirmed, this will offer important insight into the accretion origin of halo globular clusters.

In Section \ref{sec:data} we discuss the PAndAS data that are used throughout this paper. Section \ref{sec:density} contains stellar density maps of the region of interest. In Section \ref{sec:dist} we use a Bayesian technique to measure the magnitude of the tip of the red giant branch, and so estimate a distance to the South-West Cloud. Section \ref{sec:brightness} is concerned with the brightness of the South-West Cloud. We measure stellar metallicities in Section \ref{sec:metals}, discuss the implications of these measurements in Section \ref{sec:discuss}, and we conclude in Section \ref{sec:conclude}.

Throughout this paper, we assume the distance modulus to M31 to be $(m-M)_0 = 24.46 \pm 0.05$, or $779^{+19}_{-18}$ kpc \citep{conn+12}.

\section{Data}
\label{sec:data}

The photometry of M31 used in this paper was obtained as part of the PAndAS Large Program on the 3.6-metre Canada-France-Hawaii Telescope, which used the $0.96\times0.94$ square degree field of view MegaPrime camera to obtain imaging of M31, M33 and surrounds, in $g$ and $i$ bands. The full survey covers M31 and M33 out to distances of approximately 150 kpc and 50 kpc respectively, totalling more than $\sim390$~deg$^2$.

A description of the PAndAS survey can be found in \citet{mcconnachie+09}, with full details of the data reduction and a public release of the data in forthcoming publications (Ibata et al., in preparation; McConnachie et al., in preparation). To summarise, each pointing was exposed for 1350 seconds in $g$ and $i$, split into three dithered sub-exposures of 450 seconds. All observations were taken in generally excellent seeing conditions ($\la0.8\arcsec$), with a mean seeing of $0.67\arcsec$ in $g$-band and $0.60\arcsec$ in $i$-band. The median depth is $g = 26.0$~mag and $i=24.8$~mag ($5\sigma$). The data were pre-processed by CFHT staff using their \textit{Elixir} pipeline, to perform bias, flat, and fringe correction and determine the photometric zero point of the observations.

These images were then processed using a version of the CASU photometry pipeline \citep{irwin+lewis01} adapted for CFHT/MegaPrime observations. The pipeline includes re-registration, stacking, catalogue generation and object morphological classification, and creates merged $g$, $i$ products for use in subsequent analysis. The results of this process are a catalogue containing a single photometrically-calibrated $g$, $i$ entry for each detected object.

Based on curve of growth analysis, the pipeline classifies objects as noise detections, galaxies, and probable stars. For this work, we use all objects in the final catalogue that have been reliably classified as stars in both bands (aperture photometry classifications of -1 or -2 in both $g$ and $i$, which corresponds to point sources up to $2\sigma$ from the stellar locus). The CFHT instrumental magnitudes $g$ and $i$ are converted to de-reddened magnitudes $g_0$ and $i_0$ on a source-by-source basis, using the following relationships from \citet*{schlegel+98}: $g_0 = g - 3.793E(B-V)$ and $i_0 = i - 2.086E(B-V)$. 

Despite every effort to systematically cover the PAndAS survey region, holes do occur at the location of bright saturated stars, chip gaps, and a few failed CCDs. These holes have been filled with fake stars by duplicating information from nearby regions (for details, see Ibata et al., in preparation).

\subsection{Contamination model}
\label{sub:contamination}
Three sources of contamination in the PAndAS dataset can affect our study of the South-West Cloud: foreground Milky Way stars, unresolved compact background galaxies, and M31 halo stars. We account for this contamination using a model developed empirically from the PAndAS data in \citet{martin+13}. In this model, the density of contaminants $\Sigma$ at a given location $(\xi, \eta)$ and a given colour and magnitude $(g_0-i_0, i_0)$ is given by a three component exponential:

\begin{equation}
\Sigma_{(g_0-i_0,i_0)}(\xi, \eta) = \rm{exp}(\alpha_{(g_0-i_0,i_0)}\xi + \beta_{(g_0-i_0,i_0)}\eta +\gamma_{(g_0-i_0,i_0)}).
\end{equation}

The coordinates $(\xi,\eta)$ in this model, and throughout this paper, are a tangent-plane projection centred on M31. The contamination model is defined over the colour and magnitude ranges $0.2\leq (g_0-i_0)\leq3.0$ and $20 \leq i_0 \leq 24$. This model allows us to generate a colour-magnitude diagram (CMD) for contamination at any location in the PAndAS footprint. From these contamination CMDs, we can generate contamination luminosity functions and stellar densities for any region of the PAndAS survey. For full details, see \citet{martin+13}.

\section{Stellar density maps}
\label{sec:density}

\begin{figure*}
  \includegraphics[width=100mm, angle=270]{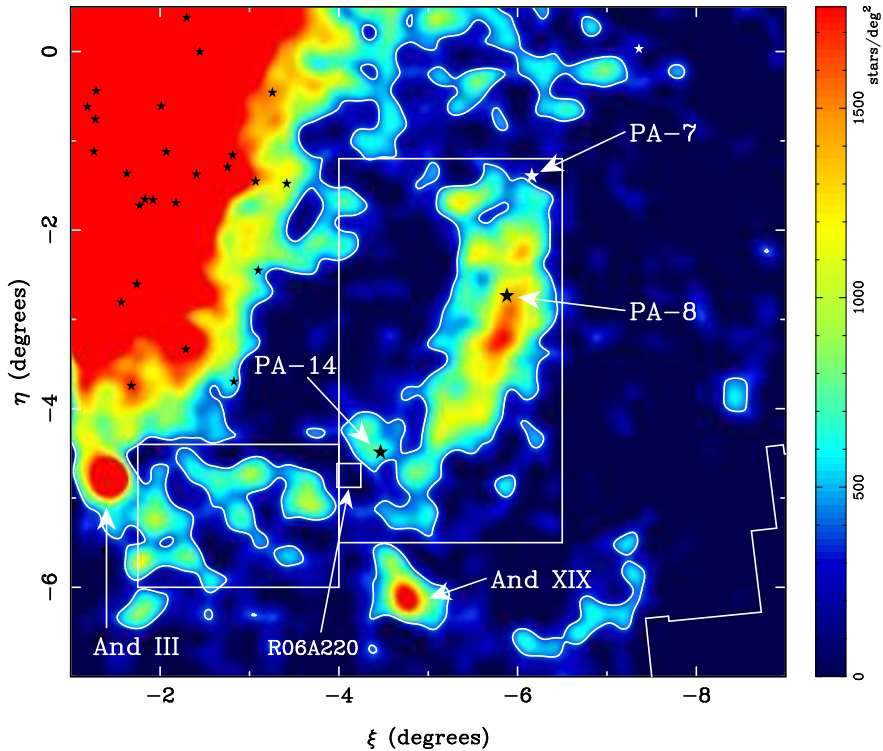}
  \caption{Zoom in on the stellar density map in the region around the South-West Cloud. Pixels are $0.025\degr \times 0.025\degr$, and have been smoothed with a Gaussian with $\sigma=0.1\degr$ dispersion. Three rectangular fields are marked: the main South-West Cloud (centre), the south-east extension (lower left, see Section \ref{sub:se_ext}), and the location of Keck DEIMOS fields R06A220 (small box; \citealt{gilbert+12}, see Section \ref{sec:discuss}). The colour gradient is plotted with linear scaling, relative to the peak density inside the box surrounding the South-West Cloud. Density contours are displayed at 25 per cent of the peak density inside the box. The positions of all known globular clusters within the region are marked with stars. Three globular clusters coincident with the South-West Cloud are highlighted: PAndAS-7 (top), PAndAS-8 (middle), and PAndAS-14 (bottom). The dwarf galaxies Andromeda III and Andromeda XIX are also marked.}
  \label{fig:partial_map}
\end{figure*}

In Figure \ref{fig:fullmap} we show a contamination-subtracted stellar density map in tangent-plane projection centred on M31 for the entire PAndAS region. Stars in this map were chosen to be consistent with metal-poor red giant branch populations at the distance of M31, using a colour-magnitude box defined by the following $(g_0-i_0,i_0)$ vertices: $(0.4,23.5)$, $(0.7, 20.9)$, $(2.3, 20.9)$, $(1.6, 23.5)$. This captures the approximate metallicity range $-2.5 < [Fe/H] < -1.0$. The map is plotted on $0.05\degr \times 0.05\degr$ pixels, and smoothed with a Gaussian with dispersion $\sigma = 0.1\degr$. 

Density maps such as these reveal a wealth of substructure, including dwarf galaxies, stellar streams, shells, and other more diffuse features. The focus of this paper, the South-West Cloud, is marked in Figure \ref{fig:fullmap} with a box. Determining the exact extent of features such as these is difficult -- no combination of magnitude or metallicity cuts unambiguously reveals where they begin and end.

Typically, some sort of radial density profile is assumed when determining whether a given star is a candidate member of a dwarf spheroidal (see for example \citealt{richardson+11}, \citealt{conn+12}, \citealt{martin+13}). For an irregular substructure such as the South-West Cloud, we instead use stellar density relative to the peak density in the region as a method for determining membership. 

In Figure \ref{fig:partial_map}, we zoom in on the region surrounding the South-West Cloud. This stellar density map has been scaled to the peak stellar density in the South-West Cloud ($\sim1767$ stars per square degree). Contours are overlaid at 25 per cent of this peak. M31 lies at the top left of this map;  two dwarf galaxies can clearly be seen: Andromeda III (below M31) and Andromeda XIX (below the South-West Cloud). All known globular clusters are marked with stars: two certainly appear to be coincident with the South-West Cloud (PAndAS-8 and PAndAS-14, middle and bottom respectively), and a third is quite probably coincident (PAndAS-7, top).

When using maps such as these to determine the true extent of the South-West Cloud, contamination subtraction is crucial. Particularly towards the north of the region near the South-West Cloud, foreground contamination makes it difficult to disentangle the Cloud from the denser regions surrounding M31 itself. Contamination subtraction, as has been performed for Figure \ref{fig:partial_map}, makes the substructure much clearer, although it can still be challenging to determine which regions should be considered connected.

We choose to define the region indicated by the box in Figure \ref{fig:partial_map} as the South-West Cloud field. This field captures most of the obvious South-West Cloud structure, although possible extension beyond this region (particularly to the south-east) suggests that our measurements of parameters such as brightness are likely to be underestimates. This is to be expected for an object that has been as heavily disrupted as the South-West Cloud.

The stellar density map in Figure \ref{fig:partial_map} has been smoothed with a $0.10\degr$ dispersion Gaussian kernel. This dispersion was chosen to give a clear view of the connected region around the peak in South-West Cloud stellar density. It is interesting to note that the globular cluster PAndAS-8 lies quite close in projection to that stellar density peak.

For the purposes of this paper, we will use the relative stellar density maps smoothed with the $0.10\degr$ dispersion Gaussian kernel as a method for determining membership probability. Stars are assigned a probability $p$ for membership in the South-West Cloud equal to the   stellar density in the pixel in which they fall. This probability is scaled relative to the peak in the South-West Cloud density, such that a star lying in the peak density pixel has a membership probability of $p=1$. In this way, the stellar density within the South-West Cloud field in Figure \ref{fig:partial_map} can also be considered as a map of South-West Cloud membership probability.

\begin{figure}
  \includegraphics[width=60mm]{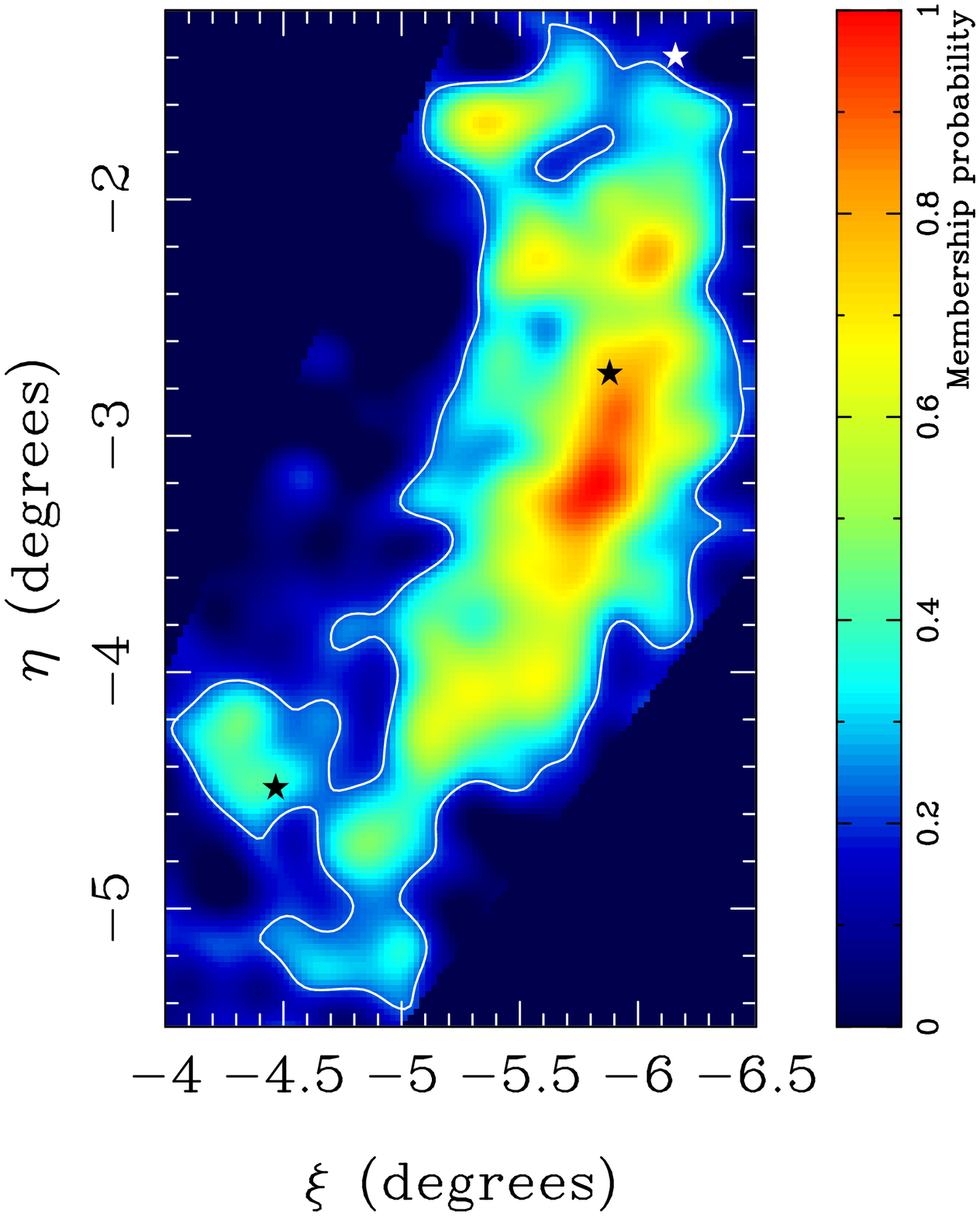}\centering
  \caption{Map of South-West Cloud membership probability. Based on smoothed, contamination-subtracted stellar density, normalised such that stars in the pixel with peak stellar density in the South-West Cloud have a membership probability of 1. Pixels are $0.025\degr \times 0.025\degr$, and have been smoothed with a Gaussian with $0.10\degr$ dispersion. The colour gradient is plotted with linear scaling, with contours at a membership probability of 0.25.}
  \label{fig:probs}
\end{figure}

With the exception of the distance determination in Section \ref{sec:dist}, membership probabilities are only used to establish a cutoff below which a star is no longer considered part of the South-West Cloud. This cutoff is taken to be a scaled stellar density of 0.25 (contours shown on Figure \ref{fig:partial_map}). We also artificially zero triangular regions to the north-east and south-west of the main South-West Cloud structure. These are unlikely to contribute significantly to measurements of distance, metallicity, or brightness, but are excluded to establish a clean boundary for the Cloud. The resulting membership probability within the South-West Cloud is show in Figure \ref{fig:probs}.

Prior to contamination subtraction, the PAndAS data for the South-West Cloud field contains 245434 stars. Applying a metal-poor red giant branch colour-magnitude cut defined by the $(g_0-i_0,i_0)$ vertices $(0.4,23.5)$, $(0.7, 20.9)$, $(2.3, 20.9)$, $(1.6, 23.5)$ reduces this to 29887 stars. After contamination subtraction, 4369 stars remain, 3216 of which lie in the region with membership probability $p\ge0.25$. We therefore have 3216 probable red giant branch members of the South-West Cloud in our sample.

Completeness fractions were determined by comparison to deep Subaru data of the Andromeda dwarf galaxy And XIV (for full details of the completeness calculation, see Martin et al., in preparation). At the tip of the South-West Cloud red giant branch, found in the next section to be $i_0 = 21.06$ mag, our data are 92 per cent complete. At the faint end of the red giant branch cut used throughout this paper, $i_0 = 23.5$ mag, the completeness fraction has fallen to 63 per cent. This fraction is essentially constant across the South-West Cloud field, although slightly worse in the southernmost quarter (57 per cent at $i_0 = 23.5$ magnitudes, $-5.5<\eta_0<-4.425$). Deeper observations would therefore heighten the contrast between South-West Cloud and the halo of M31 without drastically altering the Cloud's structure.

We note that there is structure in the Figure \ref{fig:partial_map} stellar density map extending from the south-east corner of the region we identify as the South-West Cloud, towards the dwarf galaxy AndIII. We will discuss this possible extension further in Section \ref{sub:se_ext}.

\section{Distances}
\label{sec:dist}

Distances to the South-West Cloud were estimated using the tip of the red giant branch (TRGB) as a standard candle. Here we present a summary of the method as applied to the South-West Cloud. For a full description of the Bayesian approach utilised, the reader is referred to  \citet{conn+11} and \citet{conn+12}.

To ascertain the location of the TRGB most accurately, a maximum-likelihood, model fitting approach is employed. The luminosity function of the South-West Cloud is modelled as the superposition of two luminosity functions, namely that of the Cloud's Red Giant Branch (RGB) population and that of the foreground contamination. Specifically, the RGB component is modelled as a truncated power law such that the luminosity \emph{L} varies in accordance with:

\[
  \begin{array}{l l}
L (m \ge m_{TRGB}) &= 10^{a(m - m_{TRGB})}  \\
L (m < m_{TRGB}) &= 0 

  \end{array}
\]
where $m$ is the (CFHT) $i$-band magnitude of the star in question, $m_{TRGB}$ is the TRGB magnitude and $a$ is the slope of the power law. The form of the contaminant luminosity function is ascertained by fitting a polynomial to an artificial binned luminosity function generated from contamination maps as provided in \citet{martin+13}. (Note that we differ in this treatment of the foreground contamination from that utilized in \citet{conn+11} and \citet{conn+12}, wherein representative contaminant luminosity functions are generated from adjacent reference fields.) 

Hence, to fit our model to the luminosity function of the South-West Cloud, we seek the optimum value of three parameters: the TRGB magnitude itself ($m_{TRGB}$); the slope of the RGB power law ($a$); and the fraction of contaminant stars in the luminosity function ($f$; i.e. the scaling factor of the contaminant luminosity function relative to that of the RGB of the South-West Cloud). This last parameter can be computed directly by integrating along the contaminant luminosity function to determine the total number of contaminant stars expected in the field, and hence the overall fraction of contaminant stars.   

With our model defined as described above, a Markov Chain Monte Carlo (MCMC) algorithm is used to generate probability distributions in the tip magnitude and slope model parameters. A probability distribution in the distance to the object can then be obtained by sampling from the distribution in $m_{TRGB}$. This procedure is identical to that described in Section 2 of \citet{conn+12}, and factors in the uncertainties in the absolute magnitude of the TRGB ($M^{TRGB}_i = -3.44 \pm 0.12$ mag, which includes the contribution from the uncertainty in the $\omega$Cen distance modulus; see \citealt*{B1} and \citealt{B2}) as well as in the extinction ($A_{\lambda}\pm 0.1 A_{\lambda}$). It should be noted that a uniform prior is utilised for the distance determinations in this study. 

Given the diffuse extent of the structure under study in comparison with the more compact satellite galaxies that the method has been implemented on to date, some discussion of its expected response is warranted. In particular, the present structure may extend significantly along the line of sight. If this is the case, the bright edge of the RGB would exhibit a more gentle slope rather than a sharp truncation, due to the contributions of tip stars at different line-of-sight distances. Given an extreme case of this scenario, where the structure is in effect a wall of constant density oriented edge on to the line of sight, the model fitting routine will seek a compromise tip position that best represents the data as a whole. 

\begin{figure*}
	\includegraphics[width=80mm, angle=270]{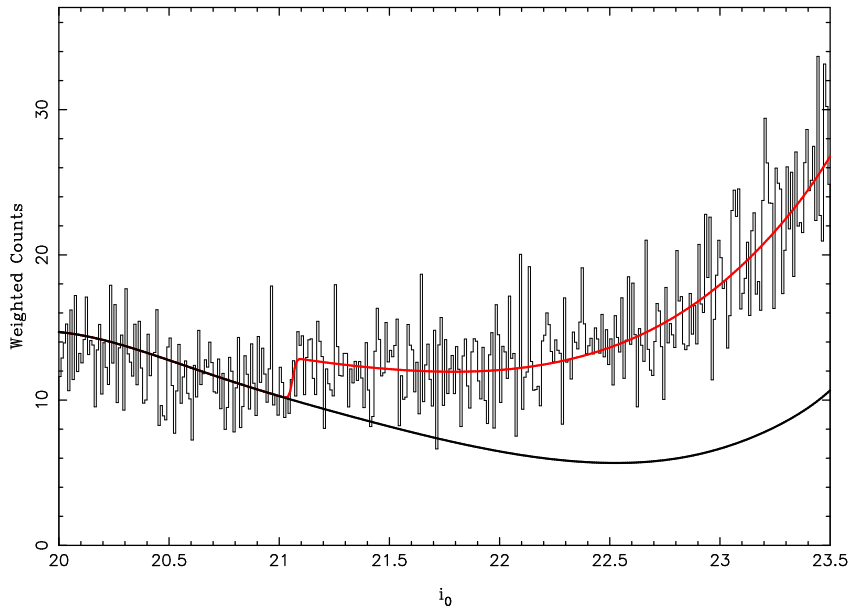}
  \caption{A histogram showing the $i_0$ luminosity function for stars in the South-West Cloud. Each star has been given a weight between 0 and 1 to reflect the density of stars in its environs relative to the maximum density in the South-West Cloud field (see Figure \ref{fig:probs}). A colour-cut has also been implemented to improve the contrast of the RGB. The best-fit model luminosity function obtained from the TRGB fitting technique is superimposed, with the South-West Cloud's RGB in red, and the foreground contamination component in black.}
  \label{fig:lum_noweighting}
\end{figure*}

\begin{figure*}
	\includegraphics[width=80mm, angle=270]{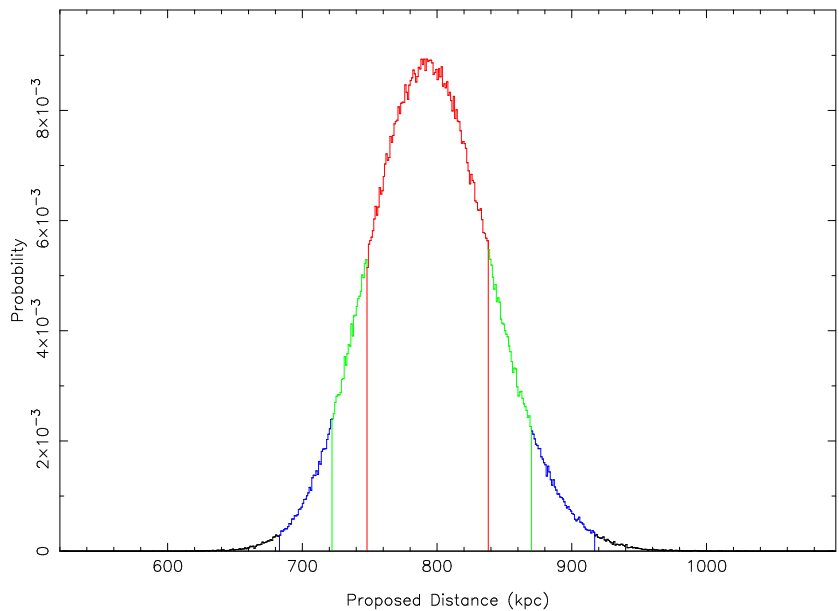}
  \caption{Probability distribution for the proposed distance in kiloparsecs, generated by evaluating the goodness of fit of the model luminosity function for 500,000 MCMC iterations. Red, green and blue denote the $1\sigma$ ($68.2 \%$), $90 \%$ and $99 \%$ credibility intervals respectively. The estimated distance to the South-West Cloud based on this distribution is $793^{+45}_{-45}$ kpc.}
  \label{fig:distance_distrib}
\end{figure*}

This said, it is important to remember that although the model used effectively assumes a structure to lie at a single distance, the method does not return a single tip magnitude or distance, but rather a probability distribution and hence the distance distribution will be substantially broadened by such a structure. Furthermore, as the structure is rotated out of this edge on orientation, it becomes increasingly possible to ascertain the distance gradient by taking measurements from sub fields at intervals along the expanding breadth of the object on the sky. In more probable scenarios however, the structure will not be uniformly dense, and hence overdensities will dominate the luminosity function, once again producing a sharp truncation to the RGB. If multiple overdensities are present at different distances, this will be evidenced by multiple peaks in the distance distribution. This is often indistinguishable from the effects of poisson noise however in the more poorly populated fields. Regardless of the scenario, the distance distributions remain representative of the field under study.      

The best-fit model obtained using this method as applied to the entire South-West Cloud field is shown superimposed on the luminosity function of the South-West Cloud in Fig. \ref{fig:lum_noweighting}. The TRGB is identified at $i_0 = 21.06^{+0.06}_{-0.06}$ mag, with $a = 0.30 \pm 0.02$ and $f = 0.75$. After factoring in an average extinction in CFHT $i$-band of $0.167$ magnitudes, a distance modulus of $(m-M)_0 =24.50^{+0.12}_{-0.13}$ is obtained, yielding an estimated distance of $793^{+45}_{-45}$ kpc. The distance distribution is presented in Fig. \ref{fig:distance_distrib} for illustration. 

In addition, the field was subsequently divided into four equal sub-fields, with distance moduli estimates from North to South as follows: $24.40^{+0.10}_{-0.25}$, $24.55^{+0.20}_{-0.12}$, $24.51^{+0.13}_{-0.13}$ and $24.34^{+0.15}_{-0.42}$, utilising extinction estimates of $0.168$, $0.194$, $0.179$ and $0.128$ magnitudes respectively. These correspond to distance estimates from North to South of $758^{+36}_{-82}$, $812^{+78}_{-43}$, $799^{+47}_{-48}$ and $737^{+54}_{-129}$ kpc. Measured distances from \citet{mackey+13} for the coincident globular clusters are consistent with the relevant sub-fields: PA-7 at $783^{+26}_{-25}$~kpc (SWC: $758^{+36}_{-82}$~kpc) and PA-8 at $824^{+27}_{-26}$~kpc (SWC: $812^{+78}_{-43}$~kpc).

\begin{figure}
\includegraphics[width=70mm]{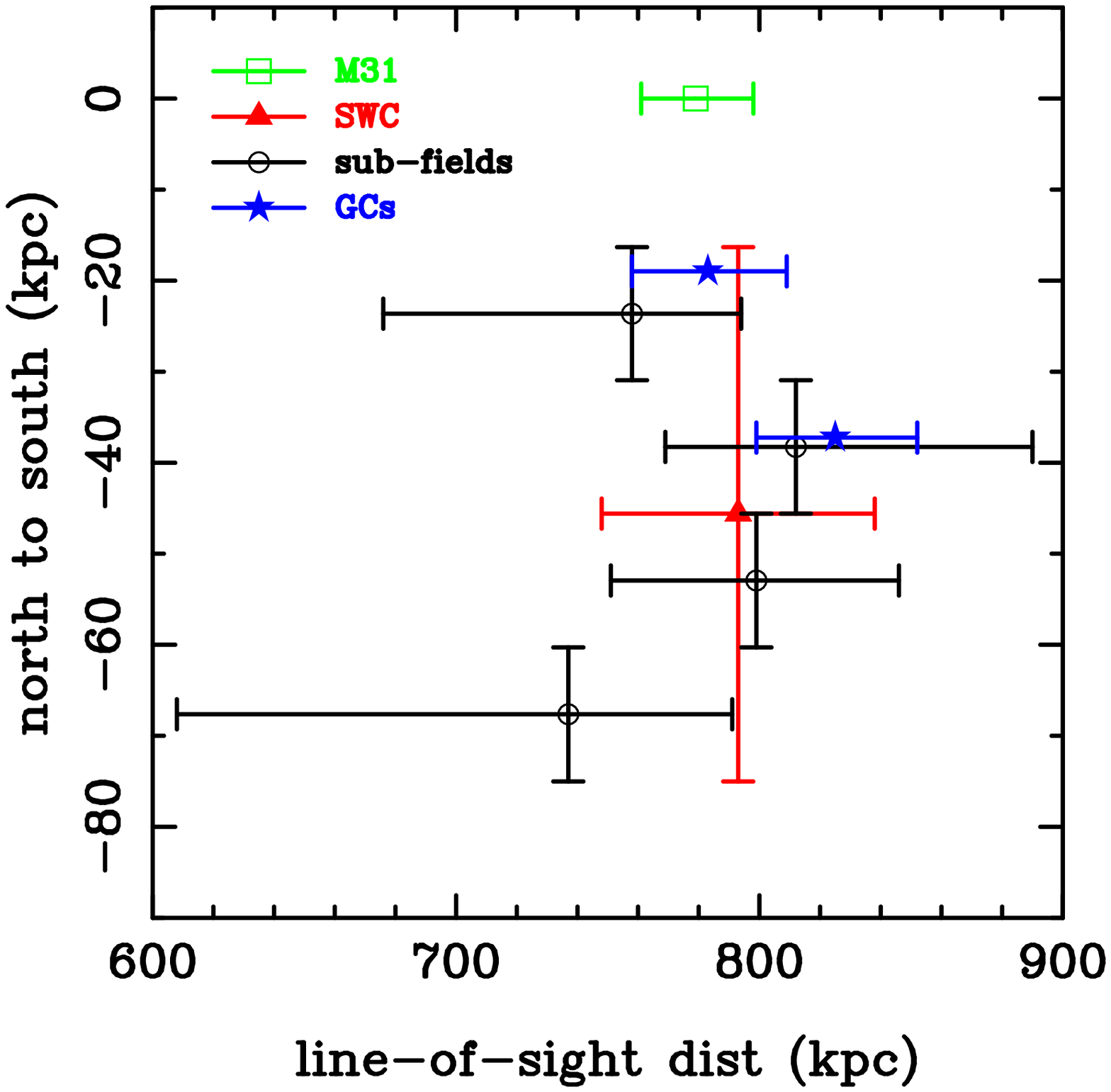}\centering
	  \caption{Distances measured to M31 \citep{conn+12}, the South-West Cloud, four equally-sized sub-fields along the length of the South-West Cloud, and the globular clusters PA-7 and PA-8 (see Section \ref{sec:dist}). The horizontal axis represents line-of-sight distance from the Earth, and the vertical axis runs north to south in projection, relative to M31. Errors in the north-south distance represent the size of the (sub-)fields, whereas errors in the line-of-sight distance are obtained formally from the tip of the red giant branch determination.}
	  \label{fig:stream}
\end{figure}

In Figure \ref{fig:stream}, we plot the distance measurements to M31, the South-West Cloud as a whole, the four sub-fields along the length of the structure, and the coincident globular clusters PA-7 and PA-8. It is tempting to interpret this as evidence that the centre of the South-West Cloud is the turning point of a stellar stream, with the central region of the field more distant than the far northern and far southern extremities. However we must be careful; although a formal measurement is obtained for the distance to the southernmost sub-field (centred on a projected distance from M31 of $\sim-65$~kpc), the signal-to-noise in the luminosity function is poor. A turning point is also inconsistent with the radial velocities of the coincident globular clusters PA-7, PA-8 and PA-14, which do not turn over. At this stage, we are unable to conclusively determine whether the South-West Cloud is indeed a stream.   

In addition to the above distances from the Milky Way, it is also possible to obtain distributions in the distances from the centre of M31. This is achieved by sampling from both the object distance distributions and that of M31 itself to obtain the object-to-M31 distance trigonometrically using the separation angle on the sky. For each field, the field centre is used for the calculation of the angular separation from the centre of the M31 disk. For the entire South-West Cloud field, an M31 distance of $86^{+27}_{-2}$ kpc is estimated. This is a good approximation for the M31 distance of the globular cluster PAndAS-8, which lies very close to the field centre. For the four subfields, the M31 distances estimated from North to South were $75^{+48}_{-2}$, $83^{+49}_{-2}$, $90^{+28}_{-2}$ and $98^{+83}_{-4}$ kpc.     

\section{Surface Brightness}
\label{sec:brightness}

The brightness of the South-West Cloud was computed by building a luminosity function (LF) from all PAndAS stars with colours and magnitudes consistent with metal-poor red giant branch populations at the distance of M31, as in Section \ref{sec:density}. To account for contamination, a matching LF was built from the contamination model described in Section \ref{sub:contamination} for the same region and colour-magnitude cut. This was subtracted from the South-West Cloud LF, and the resulting contamination-corrected LF was poisson sampled, summing the total flux to give an estimate of the South-West Cloud brightness. This process was repeated $10^4$ times, in order to determine the poisson error in the measurement. 

We perform this calculation for stars with membership probabilities $p\geq0.25$ (see Figure \ref{fig:probs}). Repeating the analysis including all stars contained within the rectangular South-West Cloud region does not significantly alter the result.

In order to correct for unresolved light, we measured the brightness of available Andromeda dwarfs within the PAndAS region, and calibrated against their $V$ band brightnesses listed in \citep{mcconnachie12}. We convert the CFHT $g$ and $i$ bands, in which the PAndAS data were taken, into the $V$ band using the following colour transform \citep{ibata+07}:

\[
 V = \left\{ 
  \begin{array}{l l}
    -0.033 + 0.714(g_0-i_0) + i_0, & \quad (g_0-i_0)<1.850,\\
    -0.480 + 0.956(g_0-i_0) + i_0, & \quad (g_0-i_0)\geq1.850,
  \end{array} \right. 
\]

The dwarfs used for calibration were: AndI, AndII, AndIII, AndV, AndIX, AndX, AndXI, AndXII, AndXIII, AndXIV, AndXV, AndXVI, AndXVII, AndXIX, AndXX, AndXXI, AndXXIII, AndXXIV, AndXXV, AndXXVI, and AndXXVII. AndXVIII was excluded as its brightness is not well known, and AndXXII was excluded as it is too close to the edge of the PAndAS region.

For each dwarf, the colour-magnitude box was shifted up or down in $i_0$ to match the tip of the red giant branch listed in \citet{conn+12}. Luminosity functions were then built using stars within each dwarf's half light radius listed in \citet{mcconnachie12}. As with the South-West Cloud, a foreground contamination LF for the same region and colour-magnitude cut was subtracted from the observed data, the corrected LF was poisson sampled, and the flux of individual stars was summed to obtain an estimate of the dwarf brightness.

Many of the dwarfs are embedded in substructure from M31, so an aperture was chosen for each dwarf to correct for M31 halo structure not accounted for in the contamination model. An aperture was deemed unnecessary for the South-West Cloud as it does not seem to overlap any substructure; this is supported by the weak dependence of our brightness estimate of the Cloud on the area chosen. For each dwarf, a LF of stars in the appropriate colour-magnitude cut in the aperture was built, and contamination-corrected as for the South-West Cloud. This LF was poisson sampled, to give an estimate of the flux in the aperture, and this value, scaled to the half light radius area for the dwarf, was subtracted from the estimate of the flux for the dwarf region.

This method recovered the known half-light brightnesses of the Andromeda dwarf spheroidals with a consistent shift of $1.9$ magnitudes. Since the half-light radius is harder to define for a disrupted object like the South-West Cloud, we repeated the above process with a much larger collection radius for each dwarf, intended to find the total brightness of each object. In addition, we re-sampled excluding various dwarfs from the analysis to verify that no single dwarf was biasing our results. The same calibration factor was found in each case, determined to be $-1.9\pm0.2$ magnitudes.

Applying this calibration to our measurement of the South-West Cloud, we find an apparent $V$-band magnitude of $12.4 \pm 0.3$ mag. Taking the distance modulus of $(m-M)_0 = 24.50^{+0.12}_{-0.13}$ obtained in the previous section, we find an absolute $V$-band magnitude of $-12.1 \pm 0.4$ mag, which places the South-West Cloud at the bright end of the Andromeda dwarf galaxy population \citet{mcconnachie12}. This corresponds to a total $V-$band luminosity of $5.6^{+2.5}_{-1.7} \times 10^6L_\odot$.

Due to the disrupted nature of the South-West Cloud, a stellar density contour was used in favour of a circular boundary for the purpose of calculating an average surface brightness. The contour used was $25$ per cent of the peak stellar density in the South-West Cloud, shown in Figure \ref{fig:probs}. This gives an average surface brightness of $31.7 \pm 0.3$ magnitudes/arcsecond$^2$.

\section{Metallicities}
\label{sec:metals}

\begin{figure*}
  \includegraphics[width=100mm]{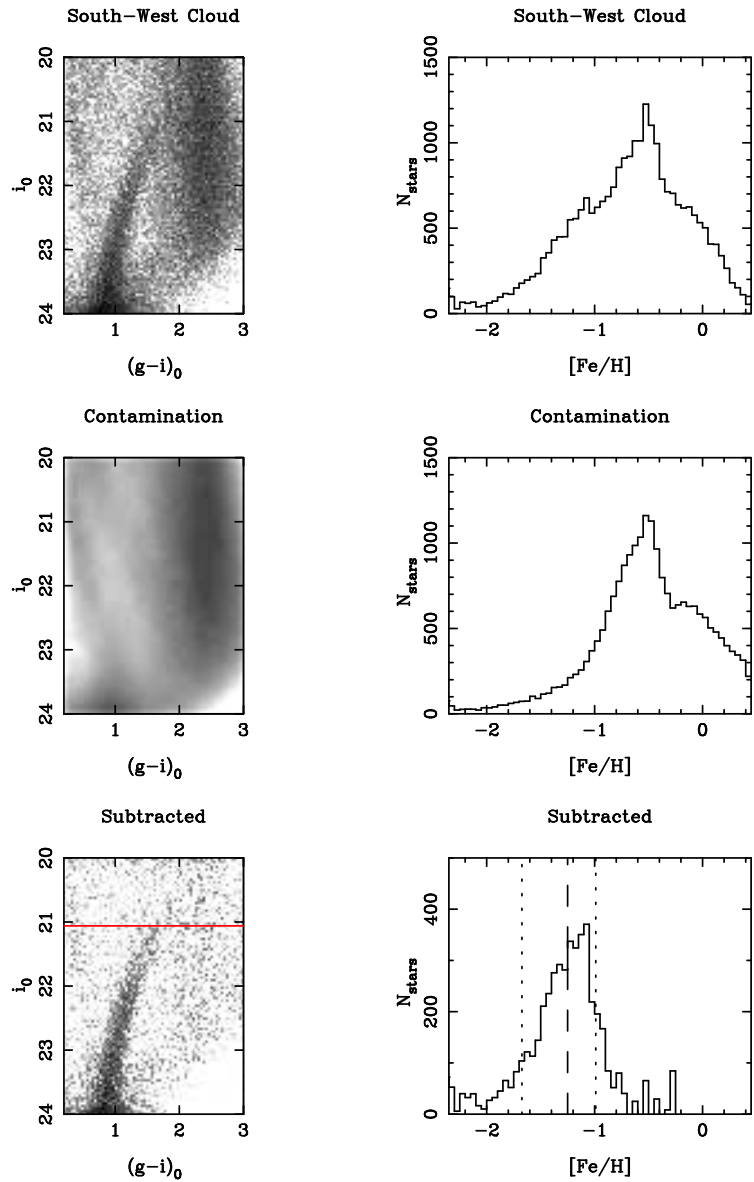}
  \caption{Left: Hess diagrams, plotted with logarithmic scaling, for the South-West Cloud (top), the contamination model for the same region (middle), and the South-West Cloud minus the contamination (bottom). Only stars falling within a pixel with density 25 per cent of the smoothed peak South-West Cloud density or greater are included (membership probability $p\geq0.25$). The tip of the red giant branch for the South-West Cloud determined in Section \ref{sec:dist} is marked with a red line in the bottom Hess diagram. Right: metallicity distribution functions generated using The Dartmouth Stellar Evolution database 2012 isochrones for a 12 Gyr population with $[\alpha/H] = 0.0$ \citep{dotter+08}.}
  \label{fig:metals_25}
\end{figure*}

\begin{figure*}
  \includegraphics[width=50mm, angle=270]{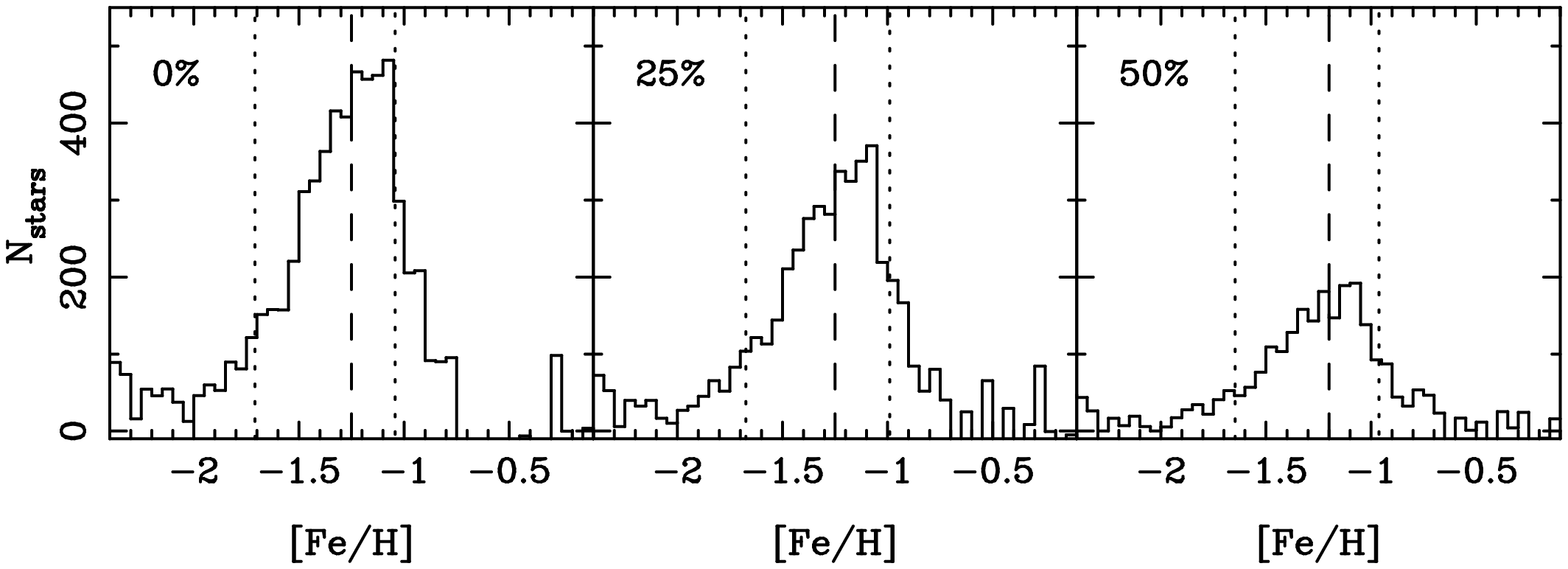}
  \caption{Metallicity distributions for stars in the entire South-West Cloud subfield ($p\geq0.0$, left), for stars that fall in regions of the smoothed map with stellar number density greater than 25 per cent of peak density ($p\geq0.25$, middle), and greater than 50 per cent of peak ($p\geq0.5$, right). In each case, contamination metallicities were subtracted to obtain final metallicity distributions. The median metallicities (dashed lines) and 68 per cent widths (dotted lines) were calculated over the metallicity range $-2.5<[Fe/H]<0.0$.}
  \label{fig:metals_cl}
\end{figure*}

We generate metallicity distribution functions (MDFs) for the South-West Cloud using stellar isochrones from the 2012 version of The Dartmouth Stellar Evolution database \citep{dotter+08}. The isochrones are available in the CFHT/MegaCam $g$, $i$ system, so no filter transformations are necessary. The de-reddened observational $(g-i)_0$ colours and $i_0$ magnitudes are compared to a grid of isochrones with 12 Gyr age, $[\alpha/Fe] = 0.0$ and metallicity $-2.5 \leq [Fe/H] \leq 0.5$ in 0.05 dex steps. The isochrones are shifted to the South-West Cloud distance modulus estimated in Section \ref{sec:dist}, $793$~kpc. Each star is assigned the metallicity of the nearest isochrone; stars further than 0.05 in colour or magnitude from any isochrone are excluded.

In Figure \ref{fig:metals_25} we illustrate the process for extracting the MDF for the South-West Cloud from the significant backgrounds. In the top row, we plot the Hess diagram for the region of the South-West Cloud field with membership probability as defined in Section \ref{sec:density} of $p\geq0.25$ (left panel). Colour-magnitude cuts were made in the data to match the complete extent of the contamination model: $0.2\leq(g-i)_0\leq3.0$, $20\leq i_0 \leq24$. The right panel shows the metallicity distribution function for this field after comparison with the stellar isochrones. We then generate a Hess diagram and MDF for an identical-sized region using the contamination model (middle row). Comparing the observational data with the contamination model, there is a clear excess in the observed MDF at $[Fe/H]\sim-1$; this is the South-West Cloud. We subtract the contamination field from the raw source field to obtain a final Hess diagram and MDF for the South-West Cloud itself (bottom row).

The median metallicity of the South-West Cloud is $[Fe/H]=-1.3\pm0.1$, taking into account errors in observed stellar magnitudes (via Monte Carlo sampling 1000 times of the PAndAS catalogue) and distance determination to the Cloud. Repeating this analysis using colour-magnitude cuts consistent with metal-poor red giant branch stars at the M31 distance, as in Section \ref{sec:density}, produces a marginally narrower metallicity spread around the same median metallicity. We emphasise that these metallicity measurements are based on an assumption of a 12 Gyr, $[\alpha/Fe] = 0.0$ population.

To determine the error-corrected metallicity spread, we treat photometric errors as metallicity differences $\Delta[Fe/H]$ (see \citealt{brasseur+11}). Monte Carlo sampling 1000 times allows us to build up an average $\Delta[Fe/H]$ distribution, the width of which is the contribution of photometric errors to the observed metallicity spread. When subtracted in quadrature from the observed metallicity spread, we obtain an intrinsic metallicity spread of $0.50\pm0.02$ dex. 

In Figure \ref{fig:metals_cl} we plot MDFs for the South-West Cloud, generated using three different cuts in stellar density (equivalently: membership probability). In the left panel, we have simply taken the entire rectangular region around the South-West Cloud ($p\geq0.0$). The middle panel shows the MDF using a 25 per cent stellar density cutoff ($p\geq0.25$, the same result is seen in Figure \ref{fig:metals_25}), and the right panel is the MDF generated using a 50 per cent cutoff ($p\geq0.5$). 

The three MDFs in Figure \ref{fig:metals_cl} are consistent with each other. This suggests that the South-West Cloud covers the majority of the rectangular field that we have selected, and that there is no significant variation in the stellar population across the field.

\subsection{South-east extension}
\label{sub:se_ext}

We have chosen in the above analysis to define the South-West Cloud as the connected region around the peak stellar density with a local membership probability of $p\geq0.25$ (equivalent to a local stellar density equal to or greater than 25 per cent of the peak stellar density). Nevertheless, there is a suggestion in the stellar density map Figure \ref{fig:partial_map} of a south-east extension of the Cloud, heading towards the dwarf galaxy And III.  

There are too few red giant branch stars, relative to the contamination, in the PAndAS data for the south-east extension to make a reliable determination of distance. To test whether this region is related to the South-West Cloud, we therefore look for a peak in the metallicity distribution function of its stars that agrees with that of the Cloud. We choose a south-east extension $(\xi, \eta)$ box that covers as much of the substructure as possible, without overlapping either AndIII or obvious M31 features (see Figure \ref{fig:partial_map}). 

\begin{figure}
 \includegraphics[width=55mm, angle=270]{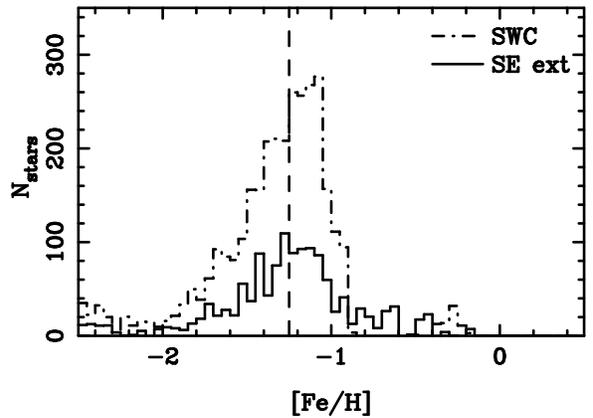}\centering
  \caption{Metallicity distributions for the South-West Cloud (dot-dashed line) and the south-east extension field (solid line) marked in Figure \ref{fig:partial_map}. Both metallicity distributions were determined for a CMD cut consistent with metal-poor red giant branch stars at the distance of M31. This cut is necessary to separate the south-east extension from the more metal-rich M31. The solid line marks the median of the south-east extension metallicity distribution, at $[Fe/H]$ = -1.2; this matches the median for the South-West Cloud itself.}
  \label{fig:se_ext_metals}
\end{figure}

In Figure \ref{fig:se_ext_metals} we plot the resulting metallicity distribution function for the south-east extension, overlaid with that of the South-West Cloud. Since the extension contains few stars relative to the contamination, we used a colour magnitude cut to isolate the signal. This cut consisted of a box defined by the $(g_0-i_0,i_0)$ vertices $(0.4,23.5)$, $(0.7, 20.9)$, $(2.3, 20.9)$, $(1.6, 23.5)$, and was used for both South-West Cloud and south-east extension fields. A radial distance of $793$~kpc was assumed for both fields.

A clear peak is seen in the metallicity distribution function for the south-east extension, with a median value of $[Fe/H] = -1.2\pm0.1$. This is formally identical to the measured median for the South-West Cloud of $[Fe/H]=-1.3\pm0.1$, suggesting that the substructure in the two fields is indeed related.

\section{Discussion}
\label{sec:discuss}

There are a number of reasons to suspect that prior to its disruption, the South-West Cloud was one of Andromeda's most luminous dwarf satellites. We can use the luminosity-metallicity relation \citep{kirby+11} to estimate the luminosity of the South-West Cloud prior to disruption:

\begin{equation}
[Fe/H] = -2.06 + (0.40 \pm 0.05) \rmn{log_{10}}\left(\frac{L_{\rmn{tot}}}{10^5L_\odot}\right)
\end{equation}

For a median South-West Cloud metallicity of $[Fe/H] = -1.3\pm0.1$, this implies a progenitor luminosity of $\sim8\times10^6L_\odot$, or an absolute magnitude of $M_V \approx -12.4$ mag. This is comparable to the brightest of the Andromeda dwarf population, And II at $M_V = -12.6$ mag and And VII at $M_V = -13.3$ mag \citep{collins+13}, and the recently discovered Lacerta I/Andromeda XXXI and Cassiopeia III/Andromeda XXXII at $M_V \sim -12$ mag \citep{martin+13b}.

The contamination-corrected region in which the South-West Cloud can be clearly seen has a measured magnitude of $M_V = -12.1$ mag. Taking the luminosity-metallicity relation at face value, the implication is that we are seeing $\sim70$-$75$ per cent of the luminosity of the progenitor object. Using the \citet{lewis+13} estimate of a stellar mass of $10^7M_\odot$ for the South-West Cloud, this implies a progenitor mass of $\sim1.5\times10^7M_\odot$.

The case for the South-West Cloud progenitor being massive is further supported by the coincidence of three globular clusters with the system: PAndAS-7, PAndAS-8 and PAndAS-14 (see Figure \ref{fig:partial_map}; \citealt{mackey+13}, \citealt{veljanoski+13}). None of the intact dwarf spheroidals in the Andromeda system are known to contain globular clusters. Indeed, only bright dwarf spheroidals in the Milky Way system are known to contain globular clusters: for example, the Sagittarius dwarf spheroidal, $M_V=-13.5\pm0.3$ mag, $> 5$ GCs, (\citealt*{ibata+95b}; \citealt{dacosta+95}; \citealt{forbes+10} and references therein) and Fornax, $M_V = -13.4\pm0.3$ mag, 5 GCs \citep{hodge61}.

Although the coincidence of globular clusters and South-West Cloud is striking, the only way to confirm association is with measurements of radial velocities. Velocities are available for all three globular clusters: PA-7 $V_r = -433 \pm 8$~km/s, PA-8 $V_r = -411\pm4$~km/s \citep{mackey+13}, and PA-14 $V_r=-363\pm9$~km/s. In \citet{gilbert+12}, the SPLASH collaboration reported a detection of a cold kinematic peak in their field R06A220. This field, consisting of 3 Keck DEIMOS masks whose location is marked in Figure \ref{fig:partial_map} with a small square, lies very close to the globular cluster PA-14. The cold kinematic peak has a measured velocity of $V_r = -373.5\pm3.0$~km/s, and a velocity dispersion of $\sigma_v = 6.1^{+2.7}_{-1.7}$~km/s. 

The velocity of this cold kinematic peak is in good agreement with that obtained for the nearby PA-14, suggesting that there is an association at the south-east corner of the Cloud. In light of this, measured velocities for the central (near PA-8) and northern (near PA-7) regions of the Cloud become particularly interesting. The globular clusters display a velocity gradient from north to south: PA-7 is moving towards the Milky Way most rapidly ($V_r = -433 \pm 8$~km/s, \citealt{mackey+13}), and PA-14 least rapidly ($V_r =-363\pm9$~km/s, J. Veljanoski priv. comm.). In combination with our distance measurements along the South-West Cloud, this offers the tantalising possibility that velocity measurements of stars in the South-West Cloud could be used to constrain the orbit of the object. 

PA-7 and PA-8 have measured metallicities of $[Fe/H] = -1.35\pm0.15$ \citealt{mackey+13}, very similar to our South-West Cloud measurement of $[Fe/H] = -1.3\pm0.1$. Although this coincidence is suggestive, by itself it does not constitute strong evidence that they are associated. Note that the metallicities for the South-West Cloud and the globular clusters were obtained using two different techniques. In the former case, individual stars were compared with theoretical isochrones for a population with 12 Gyr age at the measured South-West Cloud distance, whereas \citet{mackey+13} compared the PA-7 and PA-8 colour magnitude diagrams with fiducial globular clusters in the Milky Way.

\citet{lewis+13} presented a detection of a spur of \mbox{H\,{\sc i}} gas coincident with the South-West Cloud. The measured radial velocity of this gas is broadly consistent with the velocities of the globular clusters PA-7 and PA-8, however the error on the \mbox{H\,{\sc i}} velocity is quite large ($\sim200$km/s). If the \mbox{H\,{\sc i}} gas is associated with the South-West Cloud, this implies that any encounter between it and M31 must have been quite recent; in any extended interaction, the gas would be ram-pressure stripped away. This is broadly consistent with our determination from the luminosity-metallicity relation that the current South-West Cloud represents $\sim70$-$75$ per cent of the luminosity (or mass) of the progenitor object. More precise measurements of the velocities of the \mbox{H\,{\sc i}} gas and the South-West Cloud are required to test this spatial correlation.

We note that although the South-West Cloud lies in close proximity to the dwarf galaxies And III and And XIX (see Figure \ref{fig:partial_map}), it is probably not associated with either. And XIX is located at a formally similar radial distance to the South-West cloud ($821^{+32}_{-148}$~kpc, \citealt{conn+12}), but its measured metallicity of $[Fe/H]=-1.9\pm0.1$ \citep{mcconnachie12} is significantly different to the Cloud's $[Fe/H]=-1.3\pm0.1$. And XIX also has a measured line-of-sight velocity ($V_r = -111.6^{+1.6}_{-1.7}$~km/s, \citealt{collins+13}) that differs significantly from any of the South-West Cloud coincident globular clusters or the R06A220 cold kinematic peak ($V_r \sim -360$ to $-430$~km/s).

And III is located at $723^{+18}_{-23}$~kpc, which is inconsistent with the bulk of the South-West Cloud. However, this distance does agree with our (low signal-to-noise) measurement of the southernmost South-West Cloud field, and And III does appear to lie on the projected path of the Cloud towards the south-east. A line of sight velocity of $V_r = -344.3\pm1.7$~km/s \citep{collins+13} also lies tantalisingly on the trend in coincident globular cluster velocities. However it is difficult to imagine a scenario that would strip away an envelope of metal-rich stars (the South-West Cloud, $[Fe/H] = -1.3\pm0.1$), leaving behind a metal-poor core (And III, $[Fe/H]=-1.78\pm0.04$). 

\section{Conclusions}
\label{sec:conclude}

The South-West Cloud is a diffuse stellar substructure lying approximately 100~kpc from the centre of M31. Discovered in the Pan-Andromeda Archaeological Survey (PAndAS; \citealt{mcconnachie+09}), we present here the first detailed analysis of its stellar properties, summarised in Table \ref{tab:properties}. We identify a region $\sim50$~kpc in length at the M31 distance, over which the South-West Cloud appears to be coherent. Using a Bayesian tip of the red giant branch technique, we estimate the distance to the South-West Cloud to be $793^{+45}_{-45}$~kpc, placing it slightly behind M31 (three dimensional distance from the centre of M31: $86^{+27}_{-2}$~kpc). 

\begin{table}
 \centering
  \caption{Properties of the South-West Cloud}
  \begin{tabular}{lc}
  \hline
 \hline
Milky Way distance & $793^{+45}_{-45}$~kpc\\
M31 distance & $86^{+7}_{-1}$~kpc\\
Apparent $V$-band magnitude & $12.4\pm0.3$ mag\\
Absolute $V$-band magnitude & $-12.1\pm0.4$ mag\\
Average $V$-band surface brightness & $31.7\pm0.3$~mags/arcsec$^2$\\
Metallicity $[Fe/H]$ & $-1.3\pm0.1$\\
Metallicity spread $\Delta[Fe/H]$ & $0.50\pm0.02$\\
\hline
\end{tabular}
\label{tab:properties}
\end{table}

By comparing stars in the region to theoretical stellar isochrones for a 12 Gyr population, we obtain a metallicity for the South-West Cloud of $[Fe/H] = -1.3\pm0.1$, with a $1\sigma$ metallicity spread of $\Delta[Fe/H] = 0.50\pm0.02$. The South-West Cloud is therefore amongst the more metal-rich of M31's satellite dwarfs, which implies that prior to disruption it was also one of the most massive (comparable to And II and And VII). This conclusion is circumstantially supported by three coincident globular clusters lying along the South-West Cloud.

The South-West Cloud has a measured total brightness of $M_V = -12.1\pm0.4$ mag. The luminosity-metallicity relation for $[Fe/H]\sim-1.3$ suggests the South-West Cloud progenitor object had an approximate brightness of $M_V \sim -12.4$ mag. If we take this relationship at face value, then the Cloud contains $\sim70$-$75$ per cent of the brightness of the progenitor object, suggesting a relatively recent disruption. This is consistent with the possible association between the South-West Cloud and a spur of \mbox{H\,{\sc i}} gas \citep{lewis+13}; in a more extended interaction, the gas would have been ram pressure-stripped away.

A detection of a cold kinematic peak with radial velocity $V_r = -373.5\pm3.0$~km/s at the southern-most extent of our main South-West Cloud field \citep{gilbert+12} confirms the association between the Cloud and the nearby globular cluster PA-14 ($V_r = -363\pm9$~km/s). This lends further support to the already-favoured view that these globular clusters were accreted by M31 (\citealt{mackey+10}; \citealt{veljanoski+13}). This result would be strengthened if the globular clusters PA-7 and PA-8 could be decisively connected with the Cloud.

The three globular clusters PA-7, PA-8, and PA-14 display a north-south velocity gradient along the South-West Cloud, which would imply that the diffuse structure is part of a stream. Indeed, we confirm an extension to the south-east of the stellar density map of the South-West Cloud (see Figure \ref{fig:partial_map} and Section \ref{sub:se_ext}). Detailed kinematics of the rest of the South-West Cloud would allow us to confirm its association with the globular clusters PA-7 and PA-8, and begin mapping out the orbit of the progenitor object.

\section*{Acknowledgments}
NFB and GFL thank the Australian Research Council (ARC) for support through Discovery Project (DP110100678). GFL also gratefully acknowledges financial support through his ARC Future Fellowship (FT100100268). BM acknowledges the support of an Australian Postgraduate Award. ADM is supported by an Australian Research Fellowship from the ARC (DP1093431). We wish to thank the anonymous referee for comments that helped improve the quality of this paper.

\bsp

\label{lastpage}

\end{document}